\documentclass[]{article}

\usepackage{graphicx}
\newcommand{\ie}{\textit{i.e.}}
\newcommand{\eg}{\textit{e.g.}}

\newcommand{\rmd}{\ensuremath{\mathrm{d}}}
\newcommand{\orderof}[1]{\ensuremath{\mathcal{O}(#1)}}

\newcommand{\gae}{%
  \ensuremath{\lower 2pt \hbox{%
    $\, \buildrel {\scriptstyle >}\over {\scriptstyle \sim}\,$}%
    }%
  }
\newcommand{\lae}{%
  \ensuremath{\lower 2pt \hbox{%
    $\, \buildrel {\scriptstyle <}\over {\scriptstyle \sim}\,$}%
    }%
  }

\newcommand{\mpl}{\ensuremath{m_\textrm{Pl}}}
\newcommand{\mgut}{\ensuremath{m_\textrm{GUT}}}
\newcommand{\phie}{\ensuremath{\phi_\textrm{e}}}

\newcommand{\Vp}{\ensuremath{V^{\prime}}}
\newcommand{\Vpp}{\ensuremath{V^{\prime\prime}}}
\newcommand{\ns}{\ensuremath{n_s}}
\newcommand{\nt}{\ensuremath{n_\textrm{T}}}
\newcommand{\nsrun}{%
  \ensuremath{\frac{\mathrm{d} n_s}{\mathrm{d} \ln{k}}}%
  }
\newcommand{\Pscalar}{\ensuremath{P_\mathcal{R}}}
\newcommand{\Pscalarrt}{\ensuremath{\Pscalar^{1/2}}}
\newcommand{\Ptensor}{\ensuremath{P_T}}
\newcommand{\Ptensorrt}{\ensuremath{\Ptensor^{1/2}}}
\newcommand{\Pratio}{\ensuremath{\frac{P_T}{P_\mathcal{R}}}}
\newcommand{\rhoRH}{\ensuremath{\rho_\textrm{RH}}}

\newcommand{\refeqn}[2][eqn:]{Eqn.~(\ref{#1#2})}

\newcommand{\reffig}[2][fig:]{Figure~\ref{#1#2}}

\newcommand{\refsec}[2][sec:]{Section~\ref{#1#2}} %\S for section symbol
\newcommand{\insertfig}[1]{%
    \centering
    \includegraphics[keepaspectratio,width=1.00\columnwidth,
                     height=0.40\textheight]{#1}
}

\begin{document}
%%%%%%%%%%%%%%%%%%%%%%%%%%%%%%%%%%%%%%%%%%%%%%%%%%%%%%%%%%%%%%%%%%%%%%%%

%%%%%%%%%%%%%%%%%%%%%%%%%%%%%%%%%%%%%%%%%%%%%%%%%%%%%%%%%%%%%%%%%%%%%%%%
%######################################################################%
%#                       TITLE/ABSTRACT                               #%
%######################################################################%
%%%%%%%%%%%%%%%%%%%%%%%%%%%%%%%%%%%%%%%%%%%%%%%%%%%%%%%%%%%%%%%%%%%%%%%%

%% IJMPD FRONT MATTER ##################################################

\markboth{Freese, Kinney and Savage}
{Natural Inflation: status after WMAP 3-year data}

%\catchline{}{}{}{}{}
%
%%%%%%%%%%%%%%%%%%%%%%%%%%%%%%%%%%%%%%%%%%%%%%%%%%%%%%%%%%%%%%%%%%%%

\title{Natural Inflation: status after WMAP 3-year data}

%% IJMPD format
%\author{Katherine Freese}
%\address{
% Michigan Center for Theoretical Physics,
% Department of Physics, University of Michigan\\
% Ann Arbor, MI 48109\\
% ktfreese@umich.edu}
%
%\author{William H.\ Kinney}
%\address{
% Department of Physics, University at Buffalo, SUNY\\
% Buffalo, NY 14260\\
% whkinney@buffalo.edu}
%
%\author{Christopher Savage}
%\address{
% Michigan Center for Theoretical Physics,
% Department of Physics, University of Michigan\\
% Ann Arbor, MI 48109\\
% cmsavage@umich.edu}

%% Revtex format
%\author{Katherine Freese}
%\email[]{ktfreese@umich.edu}
%\affiliation{
%  Michigan Center for Theoretical Physics,
%  Department of Physics, University of Michigan,
%  Ann Arbor, MI 48109}
%
%\author{William H.\ Kinney}
%\email[]{whkinney@buffalo.edu}
%\affiliation{
%  Department of Physics, University at Buffalo, SUNY,
%  Buffalo, NY 14260}
%
%\author{Christopher Savage}
%\email[]{cmsavage@umich.edu}
%\affiliation{
%  Michigan Center for Theoretical Physics,
%  Department of Physics, University of Michigan,
%  Ann Arbor, MI 48109}

\author{
  \textbf{Katherine Freese\footnote{\texttt{ktfreese@umich.edu}},
          Christopher Savage\footnote{\texttt{cmsavage@umich.edu}}
          }\\[1ex]
  \textit{Michigan Center for Theoretical Physics,}\\
  \textit{Department of Physics, University of Michigan,}\\
  \textit{Ann Arbor, MI 48109}\\[2ex]
  %and\\[2ex]
  \textbf{William H.\ Kinney\footnote{\texttt{whkinney@buffalo.edu}}
          }\\[1ex]
  \textit{Department of Physics, University at Buffalo, SUNY,}\\
  \textit{Buffalo, NY 14260}
  }

%% Date
%\date{\today}
\date{}

\maketitle

%\begin{history}
%\received{Day Month Year}
%\revised{Day Month Year}
%\comby{Managing Editor}
%\end{history}

\begin{abstract}
  Inflationary cosmology, a period of accelerated expansion in the
  early universe, is being tested by Cosmic Microwave Background
  measurements.  Generic predictions of inflation have been
  shown to be correct, and in addition individual models are
  being tested.
  The model of Natural Inflation is examined in light of recent 3-year
  data from the Wilkinson Microwave Anisotropy Probe and shown to
  provide a good fit.  The inflaton potential is naturally flat due to
  shift symmetries, and in the simplest version is $ V(\phi)
  = \Lambda^4 [1 \pm \cos(N\phi/f)]$.  The model agrees with
  WMAP3 measurements as long as $f > 0.7 \mpl$ (where
  $\mpl = 1.22 \times 10^{19}$GeV) and $\Lambda \sim \mgut$.  The
  running of the scalar spectral index is shown to be small -- an order
  of magnitude below the sensitivity of WMAP3. 
  The location of the field
  in the potential when perturbations on observable scales are produced
  is examined; for $f>5\mpl$, the relevant part of the  potential is
  indistinguishable from a quadratic, yet has the advantage that the
  required flatness is well-motivated.  Depending on the value of $f$,
  the model falls into the large field ($f \ge 1.5 \mpl$) or small field
  ($f<1.5 \mpl$) classification scheme that has been applied to inflation
  models. Natural inflation provides a good fit to WMAP3 data.
\end{abstract}

%\keywords{inflation; CMB; WMAP; 98.80.Bp (PACS); 98.80.Cq (PACS).}

%% END IJMPD FRONT MATTER ##############################################

%\maketitle

\newpage

%%%%%%%%%%%%%%%%%%%%%%%%%%%%%%%%%%%%%%%%%%%%%%%%%%%%%%%%%%%%%%%%%%%%%%%%
%######################################################################%
%#                           BODY                                     #%
%######################################################################%
%%%%%%%%%%%%%%%%%%%%%%%%%%%%%%%%%%%%%%%%%%%%%%%%%%%%%%%%%%%%%%%%%%%%%%%%

%%%%%%%%%%%%%%%%%%%%%%%%%%%%%%%%%%%%%%%%%%%%%%%%%%%%%%%%%%%%%%%%%%%%%%%%
% INTRO ================================================================
\section{\label{sec:Intro} Introduction}

% r-n Plane Figure -----------------------------------------------------
\begin{figure}[tb]
  %\includegraphics[keepaspectratio,width=1.00\columnwidth,
  %                 height=0.40\textheight]{rnplane}
  \insertfig{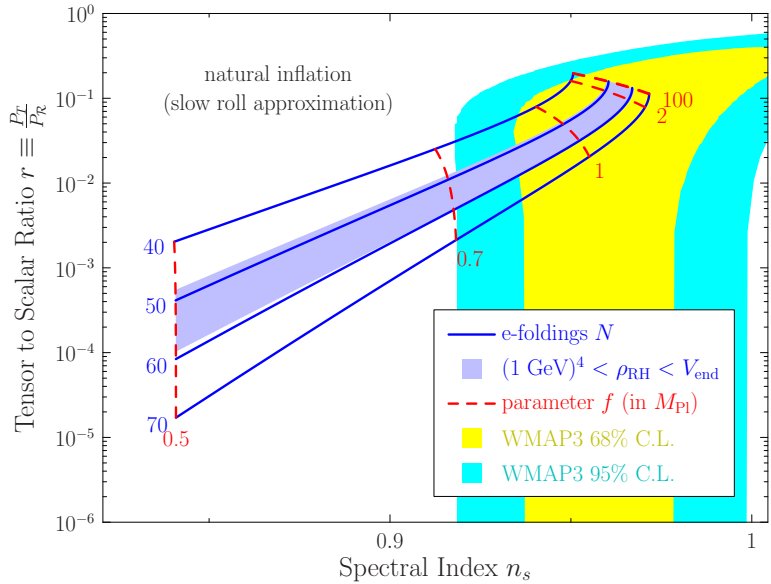}
  \caption[$r-n_s$ plane] {
    Natural inflation predictions and WMAP3 constraints in the $r$-$\ns$
    plane.  (Solid/blue) lines running from approximately the lower left
    to upper right are predictions for constant $N$ and varying $f$,
    where $N$ is the number of e-foldings prior to the end of inflation
    at which current horizon size modes were generated and $f$ is the
    width of the potential.
    The remaining (dashed/red) lines are for constant $f$ and varying
    $N$. The (light blue) band corresponds to the values of $N$ for
    standard post-inflation cosmology with $(\textrm{1 GeV})^4 <
    \rhoRH < V_\textrm{end}$.  Filled (nearly vertical) regions are the
    parameter spaces allowed by WMAP3 at 68\% and 95\% C.L.'s.  Natural
    inflation is consistent with the WMAP3 data for $f \gae 0.7\mpl$ and
    essentially all likely values of $N$.
    }
  \label{fig:rnplane}
\end{figure}
%-----------------------------------------------------------------------

Over the past five years, Cosmic Microwave Background data have
taught us a tremendous amount about the global properties of the 
universe.  We have learned about the geometry of the universe,
precise measurements of the age of the universe, the overall
content of the universe, and other properties. A ``standard 
cosmology'' with precision measurements is emerging. However,
the standard Hot Big Bang has inconsistencies, many of which
can be resolved by an accelerated period of expansion
known as inflation.  

Inflation was proposed \cite{Guth:1980zm} to solve several cosmological
puzzles: the homogeneity, isotropy, and flatness of the universe, as
well as the lack of relic monopoles.  While inflation results in an
approximately homogeneous universe, inflation models also predict small
inhomogeneities.  Observations of inhomogeneities via the cosmic
microwave background (CMB) anisotropies and structure formation are now
providing tests of inflation models.

The release of three years of data from the Wilkinson Microwave
Anisotropy Probe (WMAP3) \cite{Spergel:2006hy} satellite has generated
a great deal of excitement in the inflationary community.  
First, generic predictions of inflation
match the observations: the universe has a critical density
($\Omega=1$), the density perturbation spectrum is nearly scale
invariant, and superhorizon fluctuations are evident.  Second, current
data is beginning to differentiate between inflationary models and
already rules some of them out.  For example, quartic potentials 
and generic hybrid models do not provide a good match to the data
\cite{Spergel:2006hy,Kinney:2006qm,Alabidi:2006qa}.  
We here illustrate that the model
known as Natural Inflation is an excellent  match to current data.

Inflation models predict two types of perturbations, scalar and tensor,
which result in density and gravitational wave fluctuations,
respectively.  Each is typically characterized by a fluctuation
amplitude ($\Pscalarrt$ for scalar and $\Ptensorrt$ for tensor, with
the latter usually given in terms of the ratio $r \equiv
\Ptensor/\Pscalar$) and a spectral index ($\ns$ for scalar and
$\nt$ for tensor) describing the scale dependence of the
fluctuation amplitude.  As only two of these four degrees of
freedom are independent parameters (as discussed below), theoretical
predictions as well as data are presented in the $r$-$\ns$ plane.

Most inflation models suffer from a potential drawback: to match
various observational constraints, notably CMB anisotropy measurements
as well as the requirement of sufficient inflation,
the height of the inflaton potential must be of a much smaller scale
than that of the width, by many orders of magnitude (\ie, the potential
must be very flat).  This requirement of two very different mass scales
is what is  known as the ``fine-tuning'' problem in inflation, since
very precise couplings are required in the theory to prevent radiative
corrections from bringing the two mass scales back to the same level.
The natural inflation model (NI) uses shift symmetries to generate a
flat potential, protected from radiative corrections, in a natural way
\cite{Freese:1990rb}.  In this regard, NI is one of the best motivated
inflation models.  

One of our major results is shown in \reffig{rnplane}.
The predictions of NI are plotted in the $r$-$\ns$
plane for various parameters: the width $f$ of the potential and
number of e-foldings $N$ before the end of inflation at which present
day horizon size fluctuations were produced.  $N$ depends upon the
post-inflationary universe and is $\sim$50-60.  Also shown in the
figure are the observational constraints from WMAP's recent 3-year
data, which provides some of the tightest constraints on inflationary
models to date \cite{Spergel:2006hy}.  The primary result is that NI,
for $f \gae 0.7\mpl$, is consistent with current observational
constraints\footnote{We take $\mpl = 1.22 \times 10^{19}$ GeV.  Our
result extends upon a previous analysis of NI \cite{Freese:2004un} that
was based upon WMAP's first year data \cite{Spergel:2003cb}.}.

We emphasize two further results as well. First,
the running of the spectral index in natural inflation,
\ie\ the dependence of $\ns$ on scale is shown to be small: an
order of magnitude smaller than the sensitivity of WMAP3.  Second, we
find how far down the potential the field is at the time structure is
produced, and find that for $f > 5 \mpl$ the relevant part of the
potential is indistinguishable from a quadratic potential.  Still,
the naturalness motivation for NI renders it a superior model to a
quadratic potential as the latter typically lacks 
an explanation for its flatness.

%%%%%%%%%%%%%%%%%%%%%%%%%%%%%%%%%%%%%%%%%%%%%%%%%%%%%%%%%%%%%%%%%%%%%%%%
% INTRO ================================================================
\section{The Model of Natural Inflation\label{sec:NI}}

% MOTIVATION ===========================================================
%\subsection{\label{sec:Motivation} Motivation}

\textit{Motivation:}
To satisfy a combination of constraints on inflationary models
(sufficient inflation and CMB measurements), the
potential for the inflaton field must be very flat.  For 
models with a single slowly-rolling field, it
has been shown that the ratio of the height to the (width)$^4$ of the
potential must satisfy \cite{Adams:1990pn}
\begin{equation} \label{eqn:Vratio}
  \chi \equiv \Delta V/(\Delta \phi)^4 \le {\cal O}(10^{-6} - 10^{-8})
  \, , 
\end{equation}
where $\Delta V$ is the change in the potential $V(\phi)$ and $\Delta
\phi$ is the change in the field $\phi$ during the slowly rolling
portion of the inflationary epoch.  The small ratio of mass scales required
by \refeqn{Vratio} is known as the ``fine-tuning'' problem in inflation.

Three approaches have been taken toward this required flat potential
characterized by a small ratio of mass scales.  First, some simply say
that there are many as yet unexplained hierarchies in physics, and
inflation requires another one.  The hope is that all these
hierarchies will someday be explained.  Second, models have been
attempted where the smallness of $\chi$ is stabilized by supersymmetry.
However,  the required mass hierarchy, while stable, is
itself unexplained.  In addition, existing models have limitations.

Hence, in 1990 a third approach was proposed, Natural Inflation
\cite{Freese:1990rb}, in which the inflaton potential is flat due to
shift symmetries.  Nambu-Goldstone bosons (NGB) arise whenever a
global symmetry is spontaneously broken.  Their potential is exactly
flat due to a shift symmetry under $\phi \rightarrow \phi + \textrm{
constant}$. As long as the shift symmetry is exact, the inflaton
cannot roll and drive inflation, and hence there must be additional
explicit symmetry breaking.  Then these particles become pseudo-Nambu
Goldstone bosons (PNGBs), with ``nearly'' flat potentials, exactly as
required by inflation.  The small ratio of mass scales required by
\refeqn{Vratio} can easily be accommodated. For example, in the case
of the QCD axion, this ratio is of order $10^{-64}$.  While inflation
clearly requires different mass scales than the axion, the point is
that the physics of PNGBs can easily accommodate the required small
numbers.

The NI model was first proposed in
\cite{Freese:1990rb}.  Then, in 1993, a second paper followed which
provides a much more detailed study \cite{Adams:1992bn}.  
Many types of candidates have subsequently been explored for natural
inflation, e.g.,  \cite{Kawasaki:2000yn}$^-$\cite{Freese:1994fp}.
We focus here on the original version of NI, in which there is a single
rolling field.

% POTENTIAL ============================================================
%\subsection{\label{sec:Potential} Potential}

\textit{Potential:}
The PNGB potential resulting from explicit breaking of a shift symmetry
in single field models is 
\begin{equation} \label{eqn:potential}
  V(\phi) = \Lambda^4 [1 \pm \cos(M\phi/f)] \, .
\end{equation}
We will take the positive sign in \refeqn{potential} and
$M = 1$, so the potential, of
height $2 \Lambda^4$, has a unique minimum at $\phi = \pi f$ (the
periodicity of $\phi$ is $2 \pi f$).

For appropriately chosen values of the mass scales, \eg\ $f \sim m_{pl}$
and $\Lambda \sim \mgut \sim 10^{15}$ GeV, the PNGB field $\phi$ can
drive inflation.  This choice of parameters indeed produces the small
ratio of scale required by \refeqn{Vratio}, with $\chi \sim
(\Lambda/f)^4 \sim 10^{-13}$.  While $f \sim \mpl$ seems to be a
reasonable scale for the potential width, there is no reason to
believe that $f$ cannot be much larger than $\mpl$.  In fact, Kim,
Nilles \& Peloso \cite{Kim:2004rp} as well as the idea of N-flation
\cite{Dimopoulos:2005ac} showed that an \textit{effective} potential of
$f \gg \mpl$ can be generated from two or more axions, each with
sub-Plankian scales.  We shall thus include the possibility of
$f \gg \mpl$ is our analysis and show that these parameters can fit the
data.

% EVOLUTION ============================================================
%\subsection{\label{sec:Evolution} Evolution}

\textit{Evolution of the Inflaton Field:}
The evolution of the inflaton field is described by
\begin{equation} \label{eqn:eom}
  \ddot{\phi} + 3H\dot{\phi} + \Gamma\dot{\phi} + \Vp(\phi) = 0
  \, ,
\end{equation}
where $\Gamma$ is the decay width of the inflaton.  A sufficient
condition for inflation is the slow-roll (SR) condition $\ddot{\phi} \ll
3 H \dot{\phi}$.  The expansion of the scale factor $a$, with $H =
\dot{a}/a$, is determined by the scalar field dominated Friedmann
equation,
\begin{equation} \label{eqn:friedman}
  H^2 = \frac{8\pi}{3\mpl^2} V(\phi) .
\end{equation}
The slow roll (SR) condition implies that two conditions are met:
\begin{equation} \label{eqn:epsilonA}
  \epsilon(\phi)
    \approx \frac{\mpl^2}{16\pi}
              \left[ \frac{\Vp(\phi)}{V(\phi)} \right]^2
    =       \frac{1}{16\pi}
              \left( \frac{\mpl}{f} \right)^2
              \left[ \frac{\sin(\phi/f)}{1+\cos(\phi/f)} \right]^2 \ll 1 
\end{equation}

and
\begin{equation} \label{eqn:etaA}
  \eta(\phi)
    \approx \frac{\mpl^2}{8\pi}
              \left[ \frac{\Vpp(\phi)}{V(\phi)}
                      - \frac{1}{2} \left(
                          \frac{\Vp(\phi)}{V(\phi)} \right)^2
              \right]
    =       - \frac{1}{16\pi} \left( \frac{\mpl}{f} \right)^2 \, \ll 1.
\end{equation}
Inflation ends when the field $\phi$ reaches a value $\phie$ such that
$\epsilon(\phi) < 1$ is violated, or
\begin{equation} \label{eqn:phie}
  \cos(\phie/f) = \frac{1 - 16\pi(f/\mpl)^2}{1 + 16\pi(f/\mpl)^2} \, .
\end{equation}

More accurate results can be attained by numerically solving the
equation of motion, \refeqn{eom}, together with the Friedmann
equations.  Such calculations have been performed in
Ref.~\cite{Adams:1992bn}, where it was shown the SR analysis is
accurate to within a few percent for the $f \gae 0.5\mpl$ parameter
space we will be examining.  Thus, we are justified in using the SR
approximation in our calculations.

To test inflationary theories, present day observations must be
related to the evolution of the inflaton field during the inflationary
epoch.  A comoving scale $k$ today can be related
back to a point during inflation by finding the value of $N_k$,
the number of e-foldings before the end of inflation, at which
structures on scale $k$ were produced \cite{Lidsey:1995np}.

Under a standard post-inflation cosmology, once inflation ends, the
universe undergoes a period of reheating. Reheating can be
instantaneous or last for a prolonged period of matter-dominated
expansion.  Then reheating ends at $T <T_\textrm{RH}$, and the
universe enters its usual radiation-dominated and subsequent
matter-dominated history.  Instantaneous reheating ($\rhoRH = \rho_e$)
gives the minimum number of e-folds as one looks backwards
to the time of perturbation production, while a prolonged period of
reheating gives a larger number of e-folds.

Henceforth we will use $N$ to refer to the number of e-foldings prior
to the end of inflation that correspond to the current horizon scale.
Under the standard cosmology, the current horizon scale corresponds to
$N\sim$50-60 (smaller $N$ corresponds to smaller $\rhoRH$), with a
slight dependence on $f$.  However, if one were to consider
non-standard cosmologies \cite{Liddle:2003as}, the range of possible
$N$ would be broader.  Hence we will show results for the 
range $40 \le N \le 70$.

%%%%%%%%%%%%%%%%%%%%%%%%%%%%%%%%%%%%%%%%%%%%%%%%%%%%%%%%%%%%%%%%%%%%%%%%
% DENSITY FLUCTUATIONS =================================================
\section{\label{sec:Fluctuations} Perturbations}

As the inflaton rolls down the potential, quantum fluctuations are
generated which later give rise to galaxy formation and leave their
imprint the cosmic microwave background (CMB).  We will examine the
scalar (density) and tensor (gravitational wave) purturbations
predicted by NI and compare them with the WMAP 3 year (WMAP3) data
\cite{Spergel:2006hy}.

% SCALAR MODES =========================================================
\subsection{\label{sec:Scalar} Scalar (Density) Fluctuations}

The perturbation amplitude for the density fluctuations (scalar modes) 
produced during inflation is given by
\cite{Mukhanov:1990me,Stewart:1993bc}
\begin{equation} \label{eqn:Pscalar}
   \Pscalarrt(k) = \frac{H^2}{2\pi\dot{\phi}_k} \, .
\end{equation}
Here, $\Pscalarrt(k)$ 
denotes the perturbation amplitude when a given wavelength re-enters the
Hubble radius and
the right hand side of Eq.(\ref{eqn:Pscalar}) is to be evaluated when
the same comoving wavelength ($2\pi/k$) crosses outside the horizon
during inflation.

Normalizing to the COBE \cite{Smoot:1992td} or WMAP
\cite{Spergel:2006hy} anisotropy measurements gives $\Pscalarrt \sim
10^{-5}$.  This normalization can be used to approximately fix the
height of the potential \refeqn{potential} to be
$\Lambda \sim 10^{15}$-$10^{16}$~GeV for $f \sim \mpl$, 
yielding an inflaton mass
$m_\phi = \Lambda/f^2 \sim 10^{11}$-$10^{13}$~GeV.  Thus, a potential
height $\Lambda$ of the GUT scale and a potential width $f$ of the
Planck scale are required in NI in order to 
produce the fluctuations responsible for large scale
structure.  For $f \gg \mpl$, the potential height scales as
$\Lambda \sim (10^{-3}\mpl) \sqrt{f/\mpl}$.

% Spectral Index Figure ------------------------------------------------
\begin{figure}[tb]
  %\includegraphics[keepaspectratio,width=1.00\columnwidth,
  %                 height=0.40\textheight]{ns}
  \insertfig{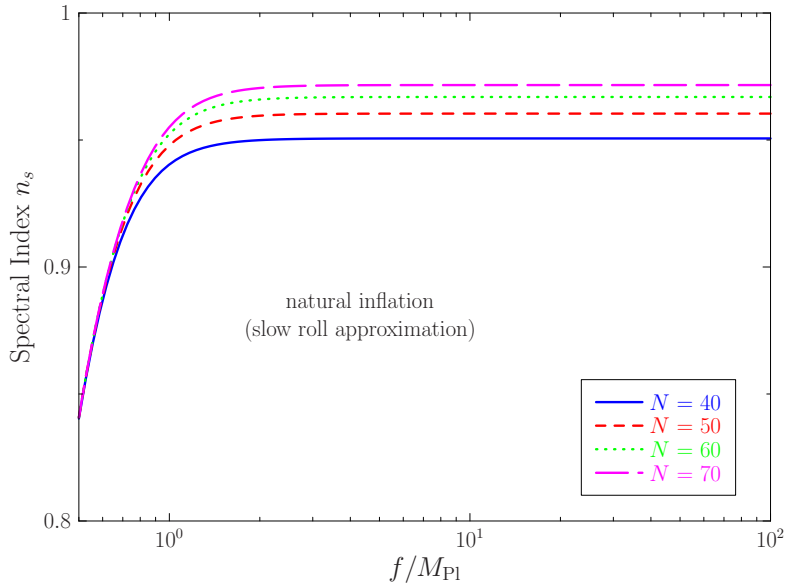}
  \caption[Spectral index $n_s$]{
    The spectral index $n_s$ is shown as a function of the potential
    width $f$ for various numbers of e-foldingss $N$ before the end
    of inflation.
    }
  \label{fig:ns}
\end{figure}
%-----------------------------------------------------------------------

The fluctuation amplitudes are, in general, scale dependent.  The
spectrum of fluctuations is characterized by the spectral index $\ns$,
\begin{equation} \label{eqn:ns}
  \ns - 1 \equiv  \frac{\rmd\Pscalar}{\rmd\ln k}
          \approx -\frac{1}{8\pi} \left( \frac{\mpl}{f} \right)^2
                  \frac{3 - \cos(\phi/f)}{1 + \cos(\phi/f)} \, .
\end{equation}
The spectral index for natural
inflation is shown in \reffig{ns}.  For small $f$, $\ns$ is essentially
independent of $N$, while for $f \gae 2\mpl$, $\ns$ has essentially no
$f$ dependence.  Analytical estimates can be obtained in these two
regimes:
\begin{equation} \label{eqn:nsA}
  \ns \approx
%  \begin{cases}
%    1 - \frac{\mpl^2}{8 \pi f^2} \, ,
%      & \textrm{for} \,\, f \lae \frac{3}{4}\mpl \\
%    1 - \frac{2}{N} \, ,
%      & \textrm{for} \,\, f \gae 2\mpl \, .
%  \end{cases}
  \left\{
  \begin{array}{ll}
    1 - \frac{\mpl^2}{8 \pi f^2} \, ,
      & \textrm{for} \,\, f \lae \frac{3}{4}\mpl \\
    1 - \frac{2}{N} \, ,
      & \textrm{for} \,\, f \gae 2\mpl \, .
  \end{array}
  \right.
\end{equation}
The WMAP 3-year data yield $\ns = 0.951_{-0.019}^{+0.015}$
($\ns = 0.987_{-0.037}^{+0.019}$ when tensor modes are included in the
fits) on the $k=0.002 {\rm Mpc}^{-1}$ scale%
\footnote{As discussed in \refsec{Running}, the running of the spectral
  index in NI is so small that the amplitude at the scale
  of the WMAP3 measurements is virtually identical to the amplitude on
  the horizon scale.}.
The WMAP3 results lead to the constraint on the width of the natural
inflation potential, $f \gae 0.7\mpl$ at 95\% C.L.

% TENSOR MODES =========================================================
\subsection{\label{sec:Tensor} Tensor (Gravitational Wave) Fluctuations}

\begin{figure}[tb]
  %\includegraphics[keepaspectratio,width=1.00\columnwidth,
  %                 height=0.40\textheight]{ratio}
  \insertfig{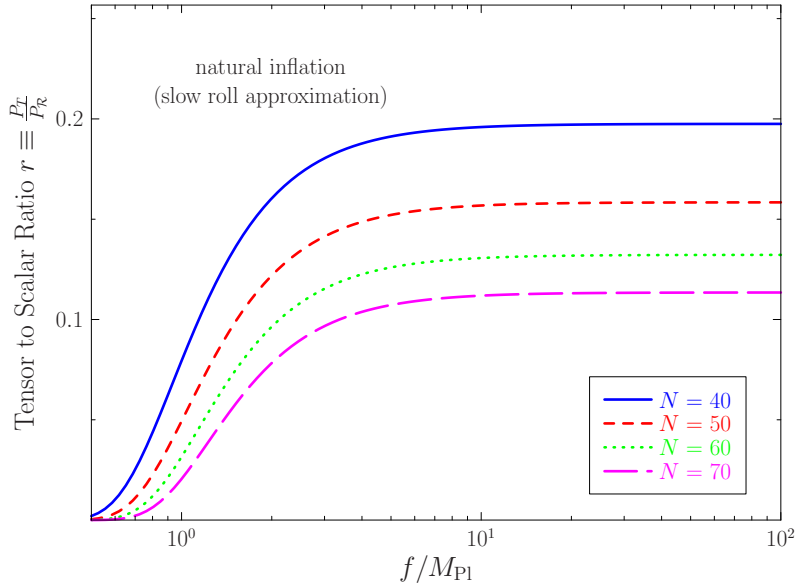}
  \caption[Tensor to scalar ratio $r$]{
    The tensor to scalar ratio $r \equiv \Pratio$ is shown as a function
    of the potential width $f$ for various numbers of e-foldingss $N$
    before the end of inflation.
    }
  \label{fig:ratio}
\end{figure}

In addition to scalar (density) perturbations, inflation also produces
tensor (gravitational wave) perturbations with amplitude
\begin{equation} \label{eqn:Ptensor}
  \Ptensorrt(k) = \frac{4H}{\sqrt{\pi}\mpl} \, .
\end{equation}
Here, we examine the tensor mode predictions of natural inflation and
compare with WMAP data.
Conventionally, the tensor amplitude is given in terms of the
tensor/scalar ratio
\begin{equation} \label{eqn:Pratio}
  r \equiv \Pratio = 16 \epsilon \, ,
\end{equation}
which is shown in \reffig{ratio} for NI.  For small $f$,
$r$ rapidly becomes negligible, while $f \to \frac{8}{N}$ for
$f \gg \mpl$.  In all cases, $r \lae 0.2$, well below the WMAP limit of
$r < 0.55$ (95\% C.L., no running).

% Tensor/Scalar Ratio Figure -------------------------------------------
%-----------------------------------------------------------------------

As mentioned in the introduction, 
in principle, there are four parameters describing scalar and tensor
fluctuations: the amplitude and spectra of both components, with the
latter characterized by the spectral indices $\ns$ and $\nt$
(we are ignoring any running here).  The amplitude of the scalar
perturbations is normalized by the height of the potential (the energy
density $\Lambda^4$).  The tensor spectral index $\nt$ is not
an independent parameter since it is related to the tensor/scalar ratio
$r$ by the inflationary consistency condition $r = -8 \nt$.
The remaining free parameters are the spectral index $\ns$ of the scalar
density fluctuations, and the tensor amplitude (given by $r$).  

Hence, a useful parameter space for plotting the model predictions
versus observational constraints is on the $r$-$\ns$ plane
\cite{Dodelson:1997hr,Kinney:1998md}.  Natural inflation generically
predicts a tensor amplitude well below the detection sensitivity of
current measurements such as WMAP. However, the situation will improve
markedly in future experiments with greater sensitivity such as QUIET 
\cite{winstein} and PLANCK \cite{unknown:2006uk}, as well as
proposed experiments such as CMBPOL \cite{Bock:2006yf}.

In \reffig{rnplane}, we show the predictions of natural inflation for
various choices of the number of e-folds $N$ and the mass scale $f$,
together with the WMAP3 observational constraints.  
For a given $N$, a fixed point is reached for $f \gg m_{pl}$; that is,
$r$ and $\ns$ become essentially independent of $f$ for any
$f \gae 10\mpl$.  This is apparent from the $f=10\mpl$ and $f=100\mpl$
lines in the figure, which are both shown, but are indistinguishable.
As seen in the figure, $f \lae 0.7\mpl$ is excluded.  However,
$f \gae 0.8\mpl$ falls well into the WMAP3 allowed region and is thus
consistent with the WMAP3 data.

%%%%%%%%%%%%%%%%%%%%%%%%%%%%%%%%%%%%%%%%%%%%%%%%%%%%%%%%%%%%%%%%%%%%%%%%
% RUNNING ==============================================================
\section{\label{sec:Running} Running of the Spectral Index}

% Spectral Index Running Figure ----------------------------------------
\begin{figure}[tb]
  %\includegraphics[keepaspectratio,width=1.00\columnwidth,
  %                 height=0.40\textheight]{nsrun}
  \insertfig{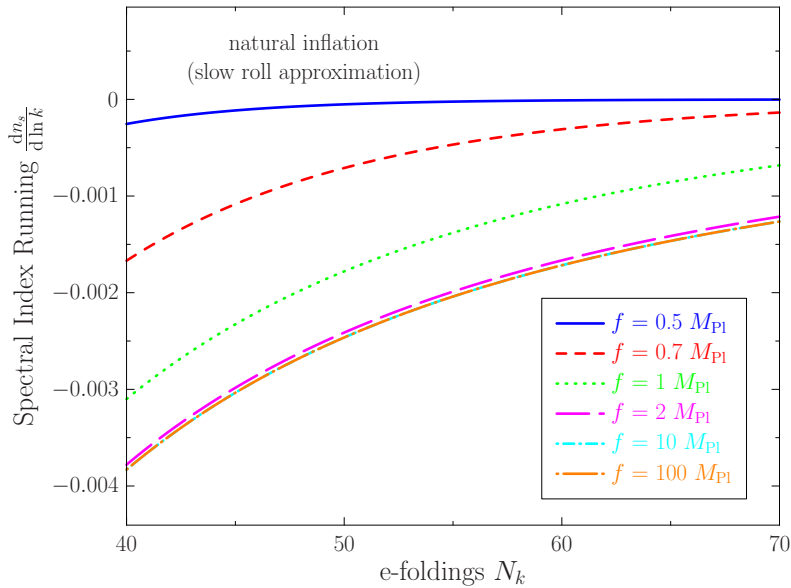}
  \caption[Spectral index running $\nsrun$]{
    The spectral index running $\nsrun$ is shown as a function of the
    number of e-foldings $N_k$ before the end of inflation for several
    values of the potential width $f$ (note that larger $N_k$
    corresponds to smaller values of $k$).  %as in Eq.(\ref{eqn:Nk})).
    Results are shown using the slow roll approximation, which is
    numerically inaccurate for this parameter, and may be off by
    up to a factor of 2-3.
    }
  \label{fig:nsrun}
\end{figure}
%-----------------------------------------------------------------------

In general, $\ns$ is not constant: its variation can be characterized
by its running, $\nsrun$.  In this section, we will use the slow roll
approximation, which is  numerically inaccurate for this parameter,
and may lead to inaccuracies of a factor of 2-3.  However, our basic
result, that the predicted running is small, is unaffected.  As shown in
\reffig{nsrun}, natural inflation predicts a small, $\orderof{10^{-3}}$,
negative spectral index running.  This is negligibly small for WMAP
sensitivities and this model is essentially indistinguishable from zero
running in the WMAP analysis.  While WMAP data prefer a non-zero,
negative running of $\orderof{10^{-1}}$ when running is included in the
analysis, zero running is not excluded at 95\% C.L.  Small scale CMB
experiments such as CBI \cite{Mason:2002tm}, ACBAR \cite{Kuo:2002ua},
and VSA \cite{Dickinson:2004yr} will provide more stringent tests
of the running and hence of specific inflation models.  
If these experiments definitively detect a strong
running (\ie, excluding a zero/trivial running), natural
inflation in the form discussed here would be ruled out.

%%%%%%%%%%%%%%%%%%%%%%%%%%%%%%%%%%%%%%%%%%%%%%%%%%%%%%%%%%%%%%%%%%%%%%%%
% POTENTIAL & MODEL SPACE ==============================================
\section{\label{sec:Potential} Inflaton Potential and Inflationary Model
                Space}

In this section, we will examine the evolution of the inflaton field
$\phi$ along the potential.  We will show that 
the location on the potential at which the
final $\sim$60 e-foldings of inflation occurs depends on the width 
$f$ of the potential.  We will also show that natural inflation
can fall into either the `large field' or `small field' categorization
defined by \cite{Dodelson:1997hr}, depending again on the value of $f$.

% POTENTIAL ============================================================
%\subsection{\label{sec:Potential2} Potential}

% Combined Potential Figure --------------------------------------------
\begin{figure*}
  %\includegraphics[keepaspectratio,width=1.00\columnwidth,
  %                 height=0.40\textheight]{potential0}
  \insertfig{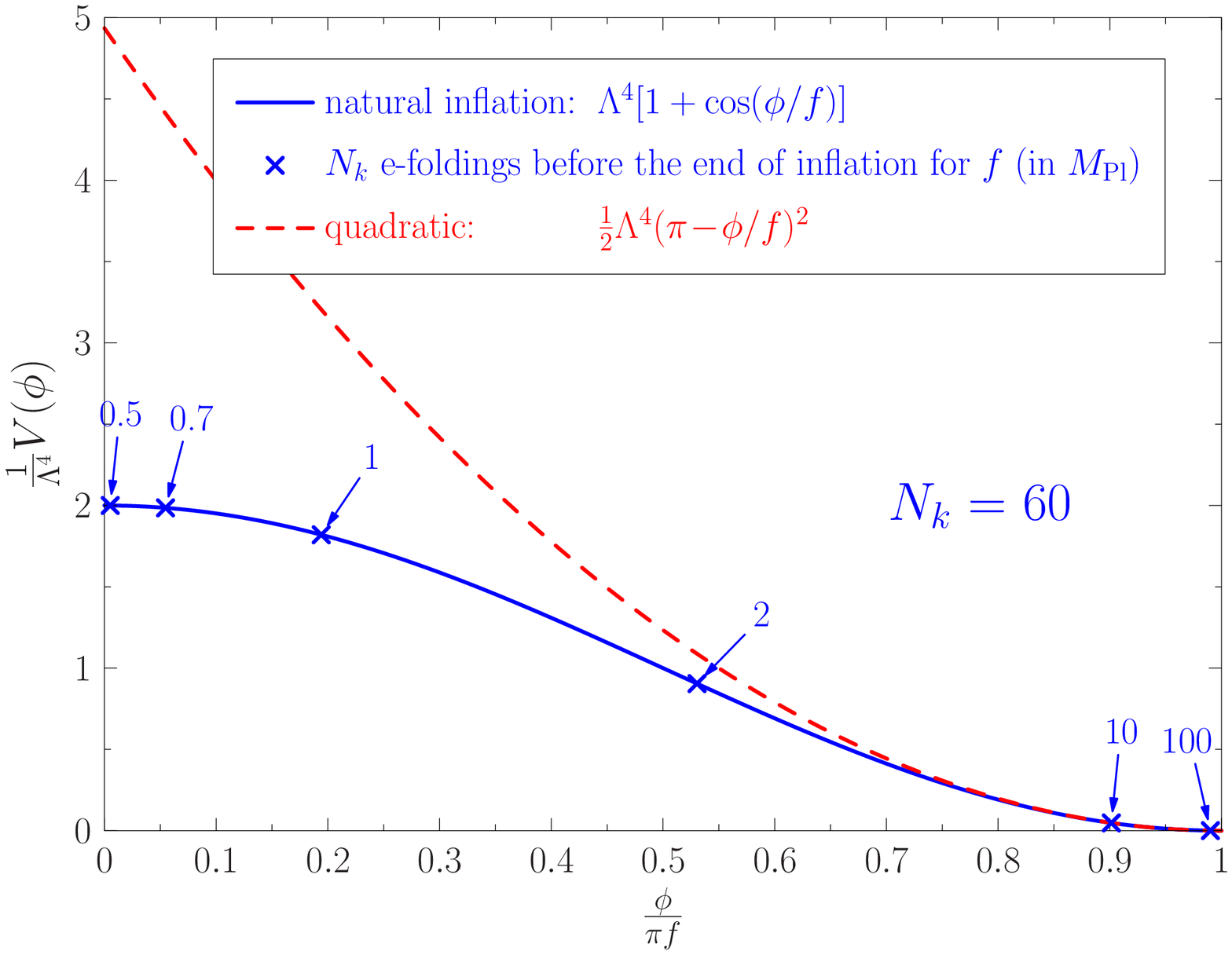}
  \hspace{\stretch{1}}
  \insertfig{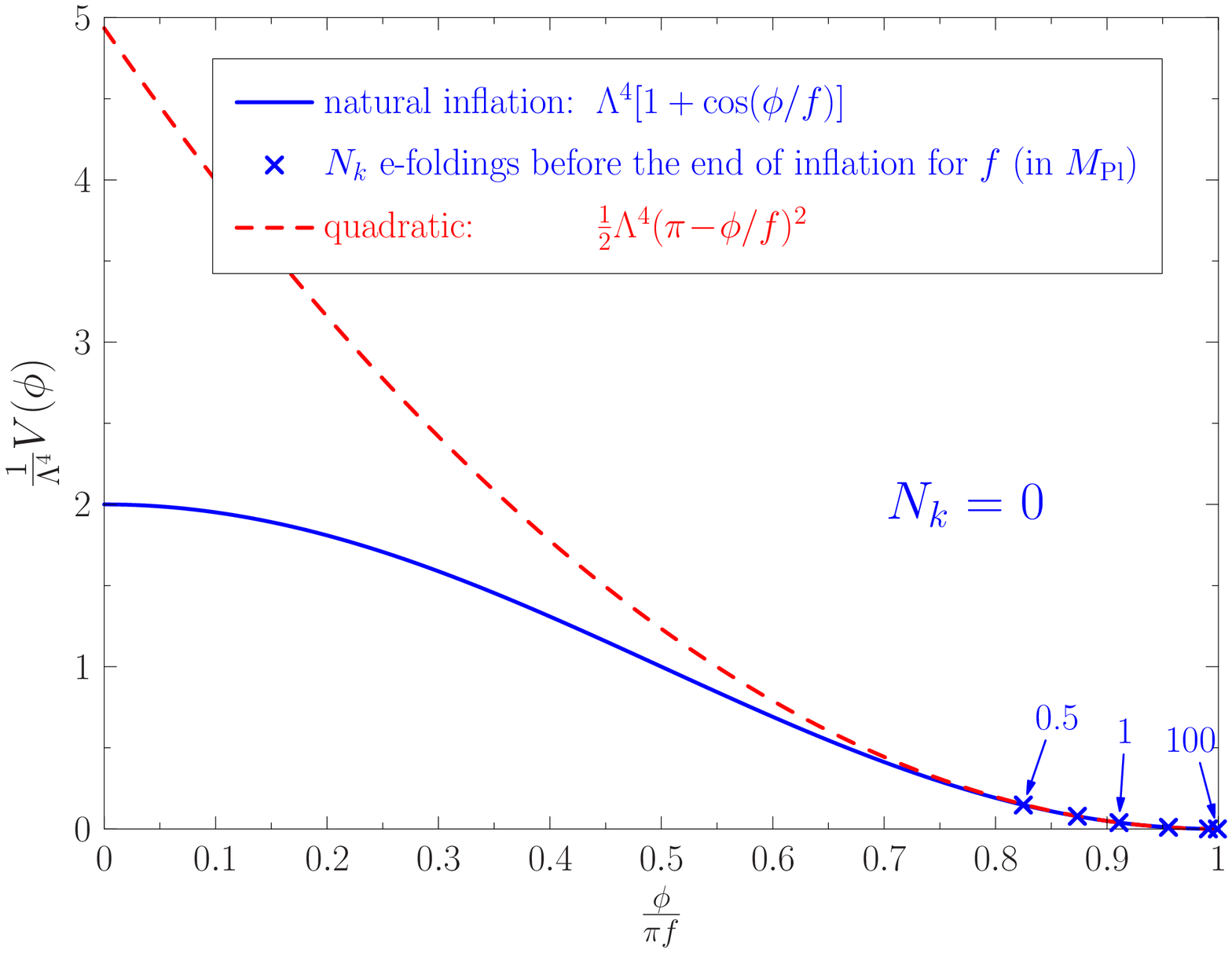}
  \caption[Inflaton potential]{
    The natural inflation potential is shown, along with a quadratic 
    expansion around the potential minimum.  Also shown are
    the positions on the potential at 60 e-foldings prior to the end
    of inflation (top panel) and at the end of inflation (bottom panel)
    for potential widths $f=(0.5,0.7,1,2,10,100) \mpl$.  For
    $f \gae 5\mpl$, the relevant portion of the potential is essentially
    quadratic during the last 60 e-foldings of inflation.
    }
  \label{fig:potential}
\end{figure*}
%-----------------------------------------------------------------------

The natural inflation potential is shown in
\reffig{potential}.  For comparison, a quadratic expansion
about the minimum at $\phi = \pi f$ is also shown.  Inflation occurs
when the field slowly rolls down the potential and ends at the point
where the field begins to move rapidly (technically, when
$\epsilon \ge 1$).  In the bottom panel of the figure, we show the
location along the potential where inflation ends ($N_k=0$) for various
values of the potential width $f$.  We see that inflation ends near the
bottom of the potential for $f>0.5 m_{pl}$.  In the top panel, the
location along the potential is shown at $N_k=60$ e-foldings prior to
the end of inflation, the approximate time when fluctuations were
produced that correspond to the current horizon -- the largest 
scales observable in the CMB. The
start of the observable portion of rolling is spread widely over the
potential.  For $f \lae 1\mpl$, current horizon modes were produced
while the field was near the top of the potential.  Conversely, for
$f \gae 3\mpl$, those modes were produced near the bottom of the
potential.  For $f \geq 5 m_{pl}$, the observationally relevant portion
of the potential is essentially a $\phi^2$ potential; note, however,
that in natural inflation this effectively power law potential is
produced via a natural mechanism.

% MODEL SPACE ==========================================================
%\subsection{\label{sec:ModelSpace} Inflationary Model Space}

% r-n Plane Figure -----------------------------------------------------
\begin{figure}[tb]
  %\includegraphics[keepaspectratio,width=1.00\columnwidth,
  %                 height=0.40\textheight]{rnplane2}
  \insertfig{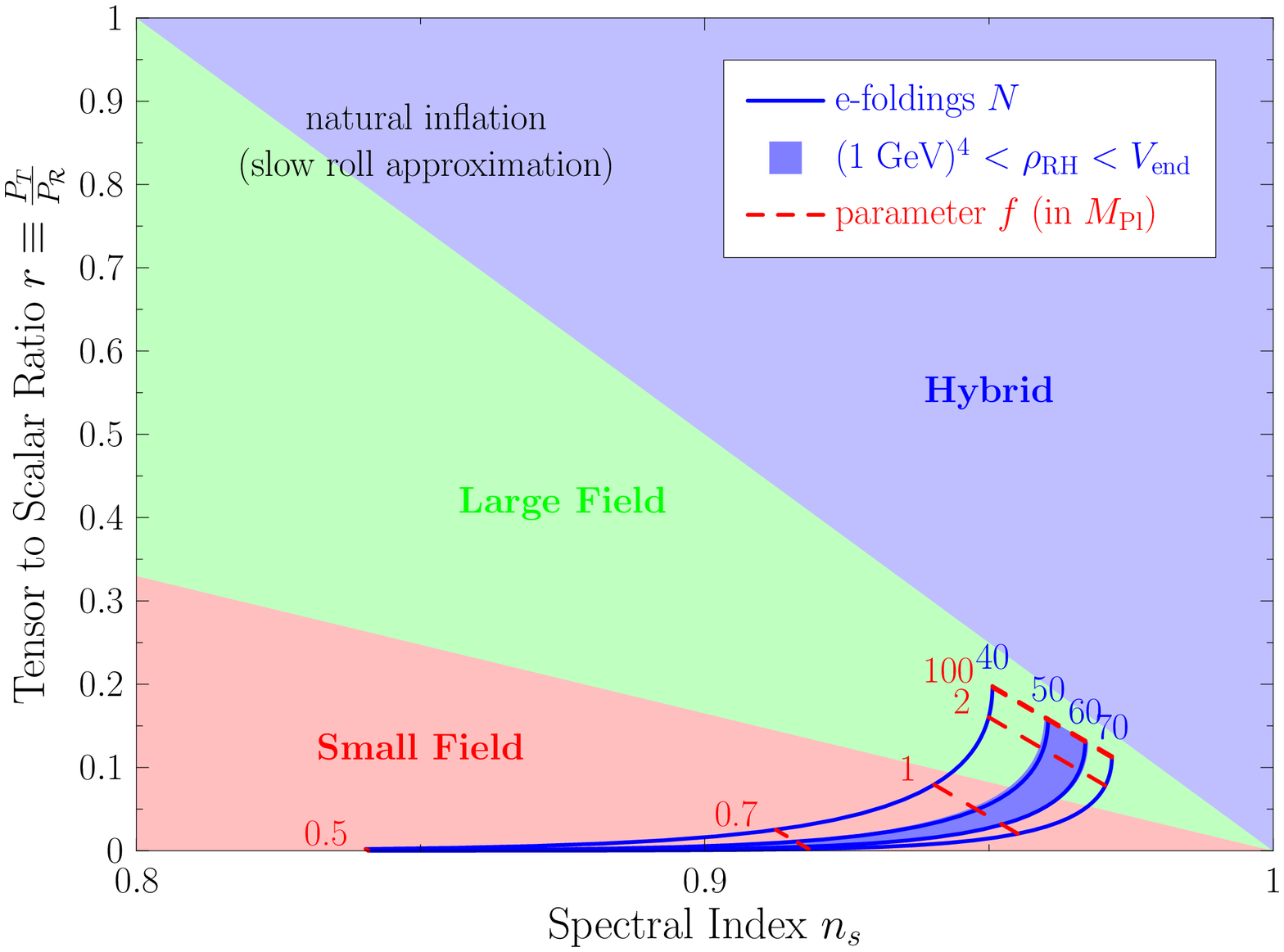}
  \caption[$r-n_s$ plane]{
    Natural inflation predictions in the $r$-$\ns$ plane (parameters
    and regions labeled as in Figure 1), as well as the regions
    classifying small field, large field, and hybrid inflation models.  
    Natural inflation falls into different classes depending on the
    potential width $f$: for $f \lae 1.5\mpl$, natural inflation can be
    classified as a small field model, while for $f \gae 1.5\mpl$,
    natural inflation can be classified as a large field model.
    }
  \label{fig:rnplane2}
\end{figure}
%-----------------------------------------------------------------------

Due to the variety of inflation models, there have been attempts to
classify models into a few groups.  
\cite{Dodelson:1997hr} have proposed a scheme with three categories:
small field, large field, and hybrid inflation models, which are easily
distinguishable in the SR approximation by the parameters $\epsilon$
and $\eta$.  To first order in slow roll, the categories have
distinct regions in the $r$-$\ns$ plane, as shown in \reffig{rnplane2}.
Also shown in the figure are the predictions for natural inflation; 
parameters are labeled as in \reffig{rnplane} (which showed the same
predictions, albeit with a logarithmic rather than linear scale).  From
\reffig{rnplane2}, it can be seen that natural inflation does not fall
into a single category, but may be either small field or large field,
depending on the potential width $f$.  This should not be surprising
from the preceding discussion of the potential.  For $f \lae 1.5\mpl$,
$\phi$ is on the upper part of the potential, where $\Vpp(\phi) < 0$,
at $N_k=60$ and, thus, falls into the small field regime.  For
$f \gae 1.5\mpl$, $\phi$ is lower down the potential, where
$\Vpp(\phi) > 0$, at $N_k=60$ and falls into the large field regime
along with power law ($V(\phi) \sim \phi^p$ for $p>1$) models.  The
WMAP3 constraints shown in \reffig{rnplane} and discussed in
\refsec{Fluctuations}, requiring $f \gae 0.7\mpl$, still allow 
natural inflation to fall into either of the small or large field
categories.

%%%%%%%%%%%%%%%%%%%%%%%%%%%%%%%%%%%%%%%%%%%%%%%%%%%%%%%%%%%%%%%%%%%%%%%%
% DISCUSSION ===========================================================
\section{\label{sec:Conclusion} Conclusion}

Remarkable advances in cosmology have taken place in the past decade
thanks to Cosmic Microwave Background experiments.  The release of the
3 year data set by the Wilkinson Microwave Anisotropy Probe is leading
to exciting times for inflationary cosmology.  Not only are generic
predictions of inflation confirmed (though there are still outstanding
theoretical issues), but indeed individual inflation models are
beginning to be tested.

Currently the natural inflation model, which is extremely
well-motivated on theoretical grounds of naturalness, is a good fit to
existing data.  For potential width $f >
0.7 \mpl$ and height $\Lambda \sim \mgut$ the model is in good
agreement with WMAP3 data.  Natural inflation predicts very little
running, an order of magnitude lower than the sensitivity of WMAP.
The location of the field in the potential while perturbations on
observable scales are produced was shown to depend on the width
$f$. Even for values $f>5 m_{pl}$ where the relevant parts of the
potential are indistinguishable from quadratic, natural inflation
provides a framework free of fine-tuning for the required potential.

There has been some confusion in the literature as to whether natural
inflation should be characterized as a 'small-scale' or 'large-scale'
model.  In Figure 8 we demonstrated that either categorization is
possible, depending on the value of $f$, and that both are in
agreement with data.

Natural inflation makes definite predictions for tensor modes, as
shown in Figure 1. Polarization measurements in the next decade have
the capability of testing these predictions and of nailing down the
right type of inflationary potentials.

%%%%%%%%%%%%%%%%%%%%%%%%%%%%%%%%%%%%%%%%%%%%%%%%%%%%%%%%%%%%%%%%%%%%%%%%
%######################################################################%
%#                      ACKNOWLEDGMENTS                               #%
%######################################################################%
%%%%%%%%%%%%%%%%%%%%%%%%%%%%%%%%%%%%%%%%%%%%%%%%%%%%%%%%%%%%%%%%%%%%%%%%

%% Acknowledgments
%\begin{acknowledgments}
\section*{Acknowledgments}
  C.S.\ and K.F.\ acknowledge the support of the DOE and the Michigan
  Center for Theoretical Physics via the University of Michigan. K.F.
thanks R. Easther and L. Verde for useful discussions.
%\end{acknowledgments}

%%%%%%%%%%%%%%%%%%%%%%%%%%%%%%%%%%%%%%%%%%%%%%%%%%%%%%%%%%%%%%%%%%%%%%%%
%######################################################################%
%#                        APPENDICES                                  #%
%######################################################################%
%%%%%%%%%%%%%%%%%%%%%%%%%%%%%%%%%%%%%%%%%%%%%%%%%%%%%%%%%%%%%%%%%%%%%%%%

%% Appendices (* for to suppress numbering for a single appendix)
%\appendix*

%%%%%%%%%%%%%%%%%%%%%%%%%%%%%%%%%%%%%%%%%%%%%%%%%%%%%%%%%%%%%%%%%%%%%%%%
%######################################################################%
%#                       BIBLIOGRAPHY                                 #%
%######################################################################%
%%%%%%%%%%%%%%%%%%%%%%%%%%%%%%%%%%%%%%%%%%%%%%%%%%%%%%%%%%%%%%%%%%%%%%%%

%%%%%%%%%%%%%%%%%%%%%%%%%%%%%%%%%%%%%%%%%%%%%%%%%%%%%%%%%%%%%%%%%%%%%%%%
\end{document}